\documentclass[acmsmall,nonacm]{acmart}

\AtBeginDocument{%
  \providecommand\BibTeX{{%
    \normalfont B\kern-0.5em{\scshape i\kern-0.25em b}\kern-0.8em\TeX}}}

\setcopyright{none}
\renewcommand\footnotetextcopyrightpermission[1]{}
\usepackage{amstext}

\usepackage{moreverb}
\usepackage[T1]{fontenc}
\usepackage[latin1]{inputenc}
\usepackage{tracefnt}
\usepackage{lmodern}
\usepackage[english]{babel}
\usepackage{textcomp}
\usepackage{inputenc}
\usepackage{courier}
\usepackage{graphicx}
\usepackage{float}
\usepackage{xcolor}
\usepackage{url}
\usepackage{listings}

\begin{document}


\title[Designing a STBP for Analyzing IQ Requirements]{Designing a Socio-Technical Business Process for Analyzing Information Quality Requirements: Experience Report}


\author{Mohamad Gharib}
\email{mohamad.gharib@unifi.it}

\affiliation{%
  \institution{DiMaI, University of Florence}
  \streetaddress{ Viale Morgagni 65}
  \city{Florence}
  \state{Italy}
  \postcode{50134}
}

\renewcommand{\shortauthors}{Gharib, M.}


\begin{abstract}

Although many BPs involve critical activities that demand high-quality information for their successful enactment, most available BP approaches focus mainly on control-flow, and either ignore the Information Quality (IQ) perspective or consider it as a mere technical issue, instead of a social and organizational one.  This leaves a BP subject to different types of social and organizational IQ vulnerabilities that may negatively impact, or even abort the BP enactment.  To tackle this problem, a Socio-Technical BP (STBP), namely a Workflow-net with Actors (WFA-net) has been developed.  WFA-net allows for capturing IQ requirements in their social and organizational context.  This paper reports on the experience gained, findings and lessons learned while developing the WFA-net.


\end{abstract}



\maketitle

\section{Introduction}

Although many BPs involve critical activities that demand high-quality information for their successful enactment, most available BP approaches focus mainly on control-flow, and either ignore the Information Quality (IQ) perspective, or consider it as a mere technical issue, instead of a social and organizational one \cite{Gharib2016rej}.  Accordingly,  a BP is subject to different types of social and organizational IQ vulnerabilities that may negatively impact, or even abort the BP enactment. Therefore, capturing IQ requirements of BPs in their social and organizational context is essential to detect and address such vulnerabilities \cite{Gharib2016rej}.

There are two main threads of research that might tackle this problem: 1- Social Business Process Management (SBPM), in which essential elements of social software are applied to the different stages of BPM \cite{schmidt2008bpm}. However, the few existing approaches do not focus on information and none of them consider IQ to the best of my knowledge. 2- Goal-based approaches, in which some techniques for capturing an execution order between goals/tasks have been proposed. 

Although some goal-based approaches are an appropriate solution for designing socio-technical systems, where the stakeholders along with their social dependencies are considered as a fundamental component of the system. They lack the formality to appropriately capture a control-flow among their goals/tasks.  On the other hand, several BP approaches offer concrete semantics for capturing control-flow among activities of BPs, but offer no semantics for capturing the social and organizational context where BPs are enacted.  

To this end, the idea for proposing a Socio-Technical BP (STBP), namely a Workflow-net with Actors (WFA-net), for modeling and analyzing IQ requirements in their social and organizational context was depending on a goal-based modeling approach for capturing the social and organizational context where a STBP is enacted, and on a Workflow modeling approach for capturing the control-flow among the activities of a STBP.

The rest of this report is organized as follows; Sect. 2 presents the case study, and Sect. 3 discusses the requirements for designing a STBP approach.  In Sect. 4, the approach is presented, and in Sect. 5 its implementation and evaluation are briefly discussed. Sect. 6 lists and discusses the findings and lessons learned, and Sect. 7 concludes the report.

\section{Case study: The Flash Crash}

The Flash Crash was a trillion-dollar stock market crash, where a stock market is a physical or virtual place where \textit{stock traders} trade securities \cite{gharib2014detecting}. \textit{Traders} can be classified into \textit{Market Makers (MMs)} that facilitate trading on specific securities, and they can trade a large number of such securities; \textit{High-Frequency Traders (HFTs)} can trade a huge number of securities with a very high-frequency; and traders who trade a few securities are called \textit{Small traders}. Usually, \textit{stock markets} analyze the trading environment and apply their Circuit Breakers (CBs) to slow or even pause trading to prevent a market crash.

\textbf{The chronology of the Flash Crash events.} The following chronology of events is based on the joint report of the Commodity Futures Trading Commission (CFTC) and the Securities and Exchange Commission (SEC) regarding the Flash Crash \cite{trading2010preliminary}:

\begin{itemize}

\item  On the 6 of May, after the U.S. stock market opened, it started trending down influenced by concerns about the debt crisis in Europe.

\item  By 2:30 PM, the value of the Dow Jones Industrial Average (DJIA) dropped down by about 2.5\%.

\item  At 2:32 PM, a big sell order of 75,000 E-mini S\&P 500 contracts (with an approximate value of \$4.1 Billion) was placed, which was mainly bought by HFTs. 

\item Between 2:41 and 2:44 PM, the price of the E-Mini contracts drop down by around  5\%.

\item  At 2:45:28 PM, the Chicago Mercantile Exchange (CME) applied its CB to pause trading on the E-Mini contracts for 5 seconds to stop a further price declines.

\item  At 2:45:33 PM, prices stabilized again when trading resumed.

\end{itemize}

\noindent\textbf{Main reasons for the Flash Crash.} Several theories have been proposed to explain this crash, including:

\begin{itemize}

\item A human error when placing the order. However, it was confirmed that the trade under suspicion of triggering the crash was not a result of a human error.

\item The loosely-coupled nature of the market and the inefficient coordination mechanisms among the trading venues concerning the work of their CBs. More specifically, trading venues need to coordinate their CBs, otherwise, HFTs can continue trading in any venue that did not apply its CB. This is exactly what happened during the Flash Crash as CME employs its CB while New York Stock Exchange (NYSE) did not.

\item Intentionally falsified orders that have been provided by some traders.  For example, some HFTs have issued flickering quotes that last for a very short time, which makes them unavailable for most traders.  Flickering quotes can be used by HFTs to drive the market prices up or down before HFTs start their actual trades. On the other hand, some MMs have issued stub quotes to fulfill their obligations concerning in facilitate trading on certain securities. Such orders are issued with prices far away from the actual market prices, i.e., they are not intended to be traded. Throughout the Flash Crash, the great majority of the trades (more than 98\%) were executed with prices about 10\% of their market values because of stub orders, i.e., stub quotes contributed significantly to the severity of the crash.  

\end{itemize}

Several key factors that contributed to this crash were due to IQ related vulnerabilities that manifested themselves in the social and organizational context of the system \cite{gharibRefsq2015}. For instance, the incompetent coordination among several markets concerning their CBs was because of relying on incomplete and inconsistent information.   Moreover, both flickering and stub quotes can be considered untrustworthy, unbelievable and/or inaccurate information; since such orders were not intended to be executed.  If markets analyze these quotes; they might be able to detect falsified ones and mitigate their harmful effect. Similarly, if markets relied on complete and consistent information for their CBs, such a crash might be avoided.

Being a real case study of a complex system that fails mainly due to IQ issues, makes the Flash Crash a  good example for illustrating the importance of capturing and addressing IQ requirements in their social and organizational context where a STBP is enacted.

\section{Identifying requirements for designing a STBP for IQ}

 Based on a thorough analysis of the Flash Crash, the STBP should be able to analyze IQ requirements based on several IQ dimensions (e.g., accuracy, believability, trustworthiness, consistency), and some of these dimensions should be analyzed in their social and organizational context. Moreover, the STBP should be able to capture control-flow as well as information-flow among its activities. Therefore, the main existing BP approaches have been surveyed to identify the best BP language that can be adopted to develop a STBP that satisfies the previously mentioned requirements. Existing BP approaches can be broadly classified into four main classes:

\textbf{Activity-driven approaches} are the most existing and adopted approaches (e.g., Petri-nets, workflow-nets, BPMN 2.0), and they mainly focus on capturing the control flow. Therefore, the emphasis of such approaches is on how a process should be enacted.

\textbf{Data-driven approaches} \cite{Muller2008} focus mainly on identifying data entities that are used by activities of the process, how data flows among such activities, and also specifying pre- and post-constraints/conditions on the use of such data.

\textbf{Event-driven approaches} \cite{VanDerAalst1999} are mainly used for modeling processes that their activities might be initiated due to some events (e.g., information modifications, time trigger, etc.). In such approaches, an event captures when an activity should be initiated.

\textbf{Role-driven modeling approaches} \cite{Saidani2006} focus on the identification of roles involved in the process along with their interactions. That is why they are used for modeling communication-based processes, i.e., the emphasis is laid on who is responsible for executing an activity. 

These approaches can model how, what, and when activities of a BP are required, yet none of them captures IQ requirements in their social and organizational context, which can be solved by adopting a goal-based approach. To this end, the idea was adopting one of the existing Activity-driven approaches that propose concepts for capturing control and information flow. Then, extending such BP approach with concepts for analyzing IQ in the social and organizational context of the system. However, this solution was not practical due to the complexity of the resulting language. Therefore, a decision was made to model the social and organizational context where the STBP is enacted and the control flow among the activities of the STBP at two different abstraction levels.  

\section{Approach for designing STBP for analyzing IQ requirements}

The process underlining the approach for designing STBP that is specialized for modeling and analyzing IQ requirements is shown in Figure \ref{fig:Approach}.  The process consists of four main phases: 1- \textit{analyzing IQ requirements,} 2- \textit{modeling IQ requirements} in their social and organizational context relying on an extended goal-based modeling language; 3- \textit{mapping,} in which leaf goals of the goal model are mapped into activities of WFA-net (STBP); and 4- \textit{analysis,} in which the correctness of the STBP is verified.  Each of these phases is described in the rest of this section:

\begin{figure*}[!t]
\centering
\includegraphics[width=  0.95 \linewidth]{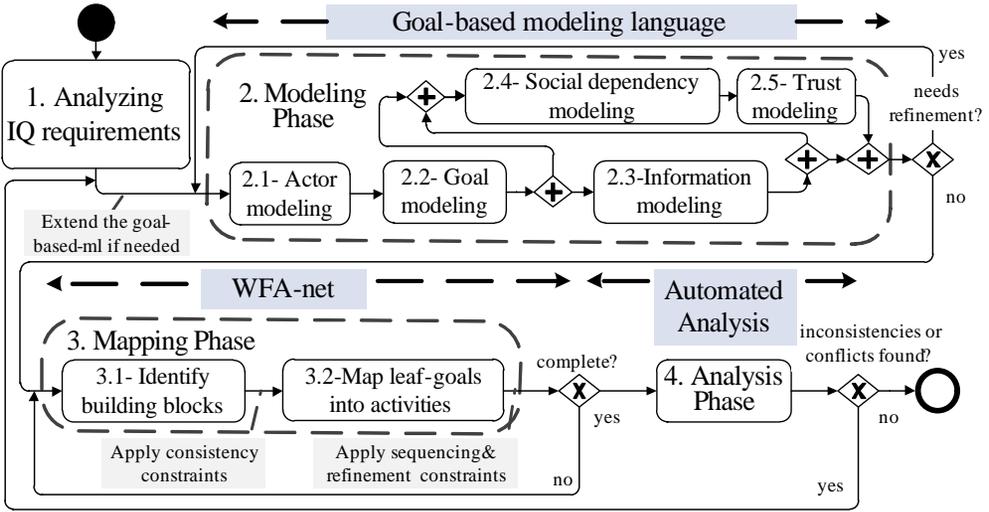}
\caption{ The process for creating the STBP for modeling and analyzing IQ requirements}
\label{fig:Approach}
\end{figure*}

\subsection{Analyzing IQ requirements}

Numerous models for analyzing IQ based on various dimensions have been proposed, yet most of them ignore the social and organizational aspects that may underlie some of these dimensions leaving the analysis incomplete for important vulnerabilities. Therefore, we proposed a multi-dimensional model for analyzing IQ requirements in their social and organizational context based on seven IQ dimensions \cite{gharibEMMSAD15}. These dimensions have been chosen based on the requirements of the stock market system, and they might need to be extended or reduced for other systems. In particular, accuracy, completeness, timeliness, and consistency are considered since they are the key IQ dimensions to tackle the vulnerabilities that led to the crash.  Additionally, believability and trustworthiness have been considered as they can be used for analyzing accuracy. Finally, accessibility has been considered to analyze information availability to perform a task. In what follows, each of the chosen dimensions is defined:

\begin{description}

\item[1. Accessibility] measures the extent to which information is available for use as well as the permissions over it that might be required for performing a specific task.

\item[2. Accuracy] measures the extent to which information is true concerning some known value. Accuracy is analyzed based on \textit{trustworthiness} and \textit{believability} since they allow for analyzing the intentional aspects underlying accuracy, i.e., dealing with situations when actors provide falsified information (e.g., flickering and stub quotes).

\item[3. Believability] measures the extent to which information is perceived as true. 

\item[4. Trustworthiness] measures the extent to which information is credible, and it is analyzed relying on the \textit{trustworthiness of the source} and \textit{trustworthiness of the provision}.   

\item[5. Completeness] measures the extent to which information is complete, and it can be analyzed based on i- \textit{value completeness} measures the extent to which the integrity of information is preserved, and ii- \textit{purpose of use completeness} measures the extent to which information is complete for performing a specific task, which cannot be analyzed properly unless the information has been considered in its organizational context. For instance, a main reason for the incompetent coordination among the CBs was because markets depend only on their local trading information to manage their CB, which can be considered incomplete for coordinating the application of their CBs with other markets. In other words, the overall system design did not consider the need of markets to coordinate their CBs, thus, markets depend on incomplete information.

\item[6. Timeliness] measures the extent to which information is valid for performing a  task.  

\item[7. Consistency] measures the extent to which all multiple records of the same information are the same across time.   \textit{Consistency} has been considered to guarantee that actors will be able to coordinate their interdependent activities, and it cannot be analyzed properly unless the information has been considered in its organizational context.  For example, information that is used for coordinating the CBs of markets also needs to be consistent among markets, since it is used for \textit{interdependent activities}.

\end{description}

\subsection{Modeling Phase}  

To capture the stakeholders' IQ requirements, the goal-based modeling framework we proposed in \cite{gharibRefsq2015,gharibEMMSAD15} has been adopted since it is the only one that proposes concepts for capturing IQ requirements.  Figure \ref{fig:gore} shows a partial diagram of the stock market system represented with our modeling language.  The language introduces primitives for modeling \textit{actors} in terms of \textit{agents} and \textit{roles} they are playing.  \textit{Goals} that represent \textit{actors'} strategic interests, and they can be refined through \textit{and/or-decomposition} into finer sub-goals. Refining a root-goal into sub-goals through \textit{and-decomposition} implies that all sub-goals need to be achieved to achieve the parent goal. While refining them through \textit{or-decomposition} implies achieving any of them achieve the root-goal.   \textit{Information} has a \textit{volatility} attribute that represents the change rate of its value. \textit{Information} can be composed of more than one part, and the \textit{part of} is used to represent such relationship.  Moreover, an \textit{actor} can be an information \textit{owner}, which enables it to control the \textit{permissions} over its usage.

 \textit{Goals} can \textit{produce}, \textit{read}, \textit{modify} and \textit{send} \textit{information}.   The \texttt{P}roduce relationship has an attribute that denotes whether such relation applies a believability check or not (shown as \texttt{B}/\texttt{NB} respectively) when producing such information.  The \texttt{R}ead relationship has three attributes, the first one is read type that can be either \texttt{O}ptional or \texttt{R}equired for goal achievement. The second captures whether the read relationship applies believability check, and the third captures the \texttt{purpose-of-use}. The \texttt{S}end relationship has an attribute that specifies the destination and another one for specifying the acceptable transmission time.

\textit{Delegation} captures the transfer of goals and permissions between actors. While \textit{provision} models information transmission between actors, and it has two attributes, the first describes the transmission time, and the second specifies the provision type, which can be normal  Provision (P) or Integrity-Preserving (IP). \textit{Trust}/\textit{distrust} (\texttt{T}/\texttt{DT}) captures the actors' expectations in one another regarding their delegated entitlements and authorities, and also for produced information (\texttt{TP}/\texttt{DP}).  

In what follows, the concepts for analyzing IQ requirements are discussed:

\begin{figure*}[!t]
\centering
\includegraphics[width=  0.99 \linewidth]{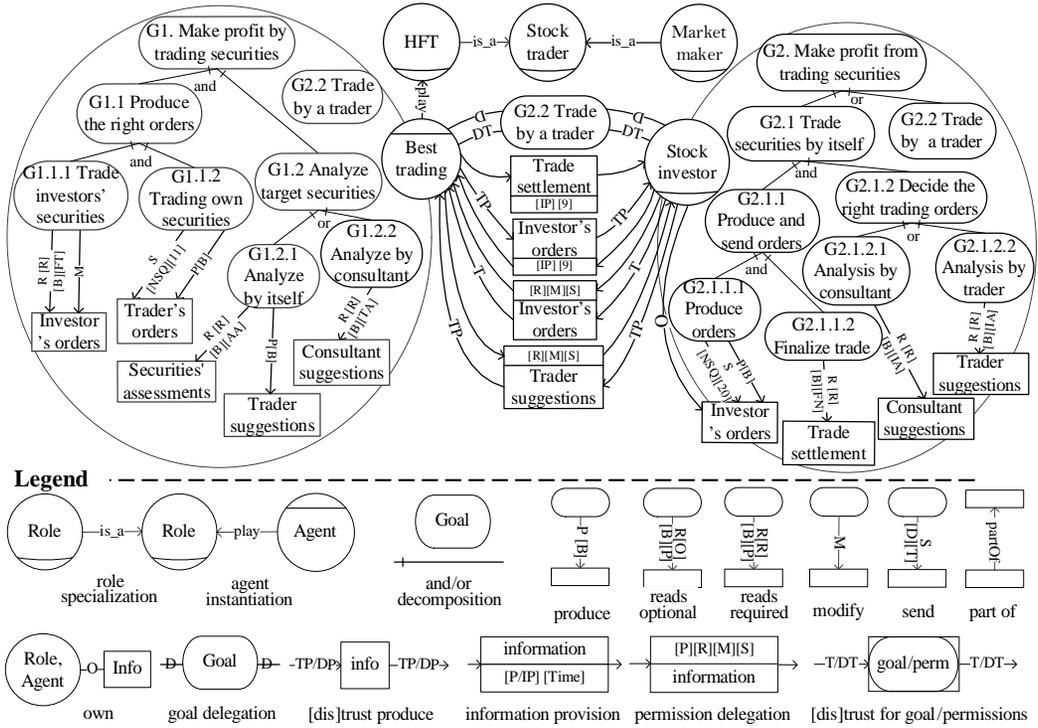}
\caption{A partial goal model concerning the stock market structure}
\label{fig:gore}
\end{figure*}  

\begin{description}

\item[1. Accessibility] is analyzed relying on information availability, and [P]roduce, [R]ead, [M]odify and [S]end permissions over such information. 

\item[2. Accuracy] is analyzed based on \textit{believability} and \textit{trustworthiness of the production}/\textit{trustworthiness of the provenance} for produced/read information.

\item[3. Believability] arises when information is being produced or read.  Therefore, believability is analyzed in produce and read relationships by examining whether such relationships apply a believability check, if yes information is considered believable.

\item[4. Trustworthiness] is analyzed based on i- \textit{trustworthiness of the source} that can be analyzed based on the \textit{trust/distrust produce} relationship; and ii- \textit{trustworthiness of the provision} that is analyzed considering the operations, which have been applied to information starting from its source until reaching its final destination, and whether such operations were authorized.

\item[5. Completeness] is analyzed based on i- \textit{value completeness,} whether the integrity of information is preserved, i.e., its value completeness is preserved if it has been provided only through \textit{IP} provision, otherwise, its value completeness is not guaranteed; and ii- \textit{purpose of use completeness,} whether information has all its sub-parts for performing a specific task, which can be analyzed with the help of the \textit{PartOf} relationship. 

\item[6. Timeliness,] only the read and send relationships are influenced by time aspects.  \textit{Read timeliness} is analyzed by comparing the currency of information with its volatility, and if its currency is smaller, information is valid.  \textit{Send timeliness} is analyzed by comparing the \textit{send time} of information and its \textit{read time} at its destination. Then,  information is valid only if the \textit{read time} is smaller than the \textit{send timeliness}.

\item[7. Consistency]  is analyzed relying on the \textit{interdependent readers} concept, which identifies actors that read the same information for the exact same \textit{purpose-of-use}. If all \textit{interdependent readers} have the same \textit{read time} for such information, it is considered consistent among them. 

\end{description}

\subsection{Mapping phase}

This section starts by describing how the semantics of WFA-net have been defined, and then discusses the developed mechanisms for mapping the goal-based model into WFA-net.

\begin{figure*}[!h]
\centering
\includegraphics[width=  0.98 \linewidth]{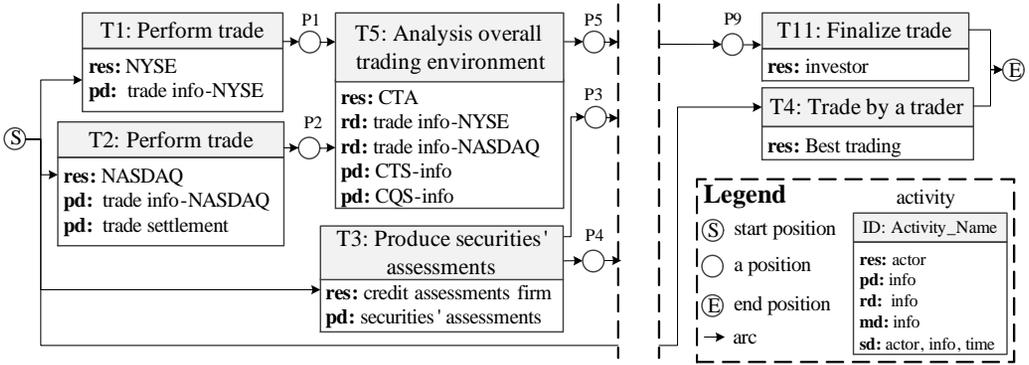}
\caption{A partial WFA-net concerning a stock investor process for trading securities}
\label{fig:WFA}
\end{figure*}

\noindent\textbf{Workflow net with Actors (WFA-net).}  To develop the WFA-net\footnote{For more information about the semantics of WFA-nets please refer to \cite{Gharib2016rej}}, social actors and their IQ requirements need to be integrated into the BP design. Thus, the WF-net has been adopted as a baseline and its semantics has been extended with the notion of social actor and IQ requirements. In particular, a WFA-net (shown in Figure \ref{fig:WFA}) is a WF-net, in which each activity $T$ is described by an actor that is responsible (res) for its achievement, sets of information that the activity produce (pd), read (rd), modify (md) and/or send (sd). 

The WFA-net configuration has been defined to capture the workflow, which consists of a \textit{marking} and an \textit{activity state} that can be evaluated either to true ($\top$) or false ($\bot$). An activity of a WFA-net might be enabled at a configuration \textit{iff} the activity is enabled at a marking (activity flow), and the activity state is evaluated to true ($\top$), which means that the information-flow and IQ requirements are achieved. An activity may fire when it is enabled, and its firing may enable a set of successor configurations. In addition, the \textit{reachability} property for WFA-net has been defined to check whether some configuration (e.g., $c^{'}$) can be reached from another one (e.g., $c$). Moreover, the firing of a single transition for WFA-net has been extended to the firing of a sequence of transitions. Finally, the \textit{soundness} property for WFA-net has been defined to verify whether the final configuration of a WFA-net is reachable from any configuration. 

\noindent\textbf{Mapping goal-based model into WFA-net.} The rules and constraints that are used to guarantee a correct mapping of goal-based models into WFA-net are illustrated in Figure \ref{fig:blocks}, and they are described as follows:

\begin{figure*}[!t]
\centering
\includegraphics[width=  0.98 \linewidth]{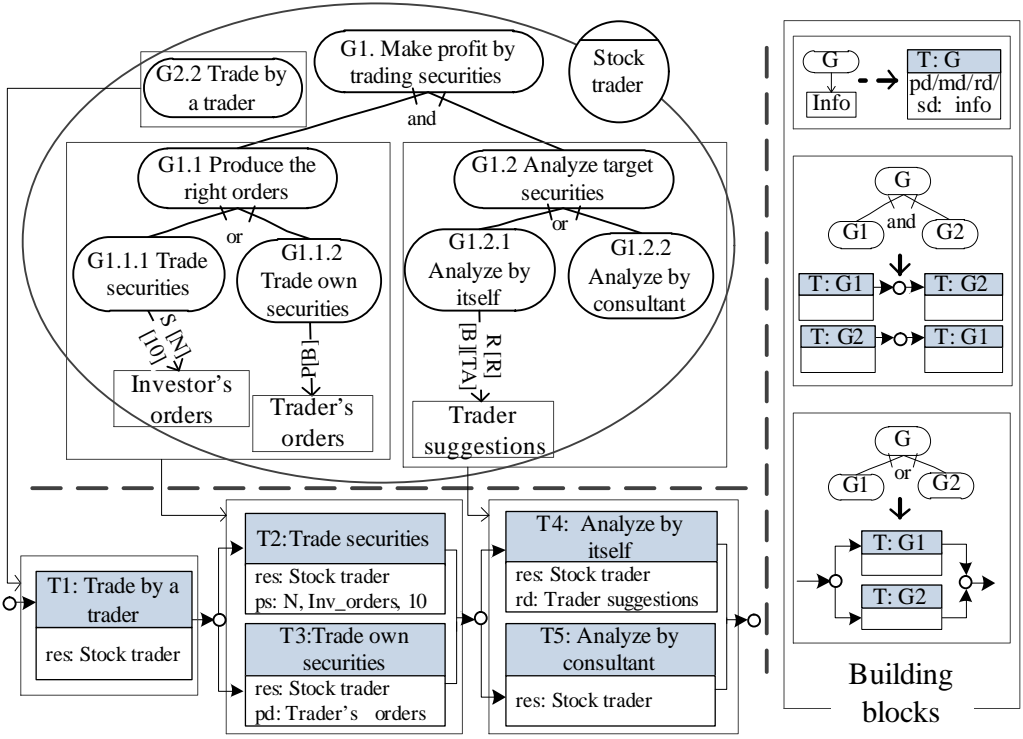}
\caption{Mapping rules and constraints}
\label{fig:blocks}
\end{figure*}

\textbf{Building blocks} are used to identify constructs of the goal-based model that can be mapped into activities of WFA-net. The main idea of identifying  \textit{building blocks}  is guaranteeing that no key organizational nor social aspects will be lost during the mapping process.  Three rules for identifying building blocks have been defined: (1) A goal that is not and/or-decomposed of or into any other goal is considered as a building block that can be mapped into an activity of WFA-net. (2)  A goal that is and-decomposed is considered as a building block in terms of all its sub-goals, which are mapped into sequencing activities. Mapping these sub-goals into sequencing activities is derived from the semantics of the and-decomposition relationship, and mapping them into sequencing activities of WFA-net implies that all of them should be achieved. (3) A goal that is or-decomposed is considered as a building block in terms of its all sub-goals, which are mapped into parallel activities of the WFA-net. Similarly, the mapping of the or-decomposed sub-goals into parallel activities is derived from the semantics of or-decomposition relationship, which implies that it is enough to achieve any one of these parallel activities in the WFA-net.

\textbf{Consistency constraints} are used to assure a correct mapping between building blocks and activities of WFA-net. Three consistency constraints have been defined: (1) No goal is allowed to be mapped unless it can be considered as a complete building block. (2) Mapping is allowed for leaf goals only, i.e., root-goals are only mapped in terms of their sub-goals. (3) No information is allowed in the WFA-net unless the goal that produces it, is already represented in the WFA-net. This enables for analyzing accessibility and other IQ dimensions.

\textbf{Sequencing constraints} are used to assure an appropriate ordering of the activities of WFA-net. Two sequencing constraints have been defined to guarantee that: (1) Activities of WFA-net should be consistent with their sequencing order in their building blocks, and (2) If an activity depends on the outcome of another one, it should appear after the later.

\textbf{Refinement constraints} are used to guarantee a correct sequencing of the activities and places of a WFA-net. Two refinement constraints have been defined: (1) No two consequent places can exist in WFA-net without an activity separating them, and (2) No two consequent activities can exist in a WFA-net without a place separating them.

\subsection{Analysis Phase}

After modeling (mapping) the desired WFA-net, its correctness should be verified. Therefore, a formal framework to underlies the modeling language has been developed based on Disjunctive Datalog formal language. This allows for converting constructs of the graphical model into formal predicates, and helps in verifying the correctness of the WFA-net\footnote{The formalization of the concepts/axioms can be found at \url{https://bit.ly/3fldgKc}}. Additionally, a set of 21 properties of the design has been defined\footnote{For more information about the properties of the design please refer to \cite{Gharib2016rej}}, which specify constraints for verifying the correctness of the mapping, control-flow, information-flow and IQ requirements of the WFA-net model.

\section{Implementation and evaluation}

The proposed approach was evaluated using a simulation method, i.e., developing a prototype implementation tool\footnote{The prototype tool is available at \url{http://goo.gl/Iy1BjR}}  and test its applicability with realistic data. More specifically, the approach has been evaluated by showing its utility and efficacy in modeling and analyzing the IQ requirements in STBPs by applying it to two realistic scenarios abstracted from the Flash Crash case study. The approach was able to models and effectively analyze (e.g., detect violations to the properties of the design) the IQ requirements of the two scenarios.

\section{Findings and lessons learned}

While developing this approach, several challenges have been faced. The main challenges are summarized along with findings and lessons learned as follows:

\noindent\textbf{A model for analyzing IQ.} There is no general agreement on the number or which dimensions should be considered for analyzing IQ. Therefore, choosing the appropriate set of IQ dimensions is not an easy task. Like many researchers, I believe that IQ dimensions should be chosen with respect to the application domain (i.e., domain-specific).  A more complex problem is how the selected IQ dimensions should be analyzed as researchers have proposed various methods/metrics for their analysis. I also believe that the analysis of IQ dimensions should be domain-specific with the help of domain experts if possible since the same dimension might be interpreted differently in different domains. For example, most of the existing models do not propose analysis for accuracy that can tackle the problem of intentionally falsified information. Therefore, accuracy has been analyzed in this approach in a way that allows for capturing the intentional aspects underlying it.

 \noindent\textbf{Modeling IQ requirements in a BP.} 1- Adopting an \textit{i}* based language \cite{gharibRefsq2015,gharibEMMSAD15,Gharib2019} for modeling IQ requirements as such languages allow for dealing with requirements considering their social and organizational aspects. Moreover, some of these modeling languages are supported by formal frameworks that allow for performing different kinds of analysis to verify the requirements model. 2- Adopting a WF-net as a base BP modeling language since it proposes a simple but powerful semantics, which can be extended to integrate almost any property of interest.

\noindent\textbf{Adopting two layers of abstraction.} Capturing the social and organizational context where the STBP is enacted, and the control flow among its activities at two different levels of abstraction offers several advantages:  1- modeling and verifying the goal-based model allows for detecting and resolving any vulnerability at the social and organizational level concerning IQ requirements before modeling the STBP; 2- relying on a verified goal model, you can model and analyze almost any  STBP without influencing the base goal model;  3- most of the properties of the design can be derived either from the semantics of the goal-based model or the WFA-net model. In WFA-net, most of the properties for the satisfaction of IQ requirements were derived from the goal-based model semantics, while most of the properties for the control and information flow were derived from the WFA-net semantics; and 4- allows for extending the goal-based modeling language to consider new IQ dimensions. Then, easily reflecting such extensions into the WFA-net.

\section{Conclusions and Future work}
   
This paper reported on experience gained while developing an STBP, namely WFA-net \cite{gharibBPMDS15,Gharib2016rej}, which integrates goal-based and WF-net approaches. The focus was laid on making the approach easy to be understood and followed; each of its steps was accompanied by a detailed description of how it was performed. This may help other scholars while dealing with IQ for BP. The proposed approach has several limitations and threats to its validity, which opens opportunities for future work. For example, the approach considers binary IQ requirement satisfaction (e.g., a requirement is satisfied or denied), it cannot deal with more than one STBP at the same time, and it considers that all IQ dimensions have the same priority/importance to the system. Finally, the approach has been applied only to the stock market domain, which threatens the generalization of the findings.

 

\end{document}